\documentclass[epj]{svjour}
\usepackage{graphics}
\begin{document}
\title{Neutron- and muon-induced background in underground physics experiments}
\author{V.~A.~Kudryavtsev\inst{1} \and L.~Pandola\inst{2} \and 
V.~Tomasello\inst{1}
}                     

\institute{
Department of Physics and Astronomy, University of Sheffield, Sheffield, United Kingdom \and
INFN, Laboratori Nazionali del Gran Sasso, Assergi, Italy}
\date{Received: date / Revised version: date}
%
\abstract{
Background induced by neutrons in deep underground laboratories is a critical issue
for all experiments looking for rare events, such as dark matter interactions or 
neutrinoless $\beta\beta$ decay. Neutrons can be produced either by natural radioactivity,
via spontaneous fission or ($\alpha$,n) reactions, or by interactions initiated
by high-energy cosmic rays. In all underground experiments, Monte Carlo simulations of neutron 
background play a crucial role for the evaluation of the total background rate and for the 
optimization of rejection strategies. 
The Monte Carlo methods that are commonly employed to evaluate neutron-induced background
and to 
optimize the experimental setup, are reviewed and discussed. Focus is given to the issue of 
reliability of Monte Carlo background estimates. 
} 
\maketitle
\section{Introduction}

The key issue for all experiments looking for rare events, such as dark matter interactions, 
neutrinoless $\beta\beta$ decay, low-energy neutrinos, etc., 
is the suppression and rejection of the 
background due to 
cosmic rays and natural radioactivity. The former is reduced by operating the experiments in 
deep underground laboratories: the rock shielding, a few km of water equivalent (w.~e.), 
allows the reduction of the flux of cosmic ray muons by a factor of $10^{5}-10^{7}$ with respect 
to the surface. Background due to natural 
radioactivity (mainly $\gamma$-rays) can be 
suppressed by means of passive shielding or rejected by specific identification tools, depending 
on the experimental design (\emph{e.g.} $\gamma$-recoil discrimination for dark matter experiments, 
single-site selection for neutrinoless $\beta\beta$ decay experiments). 

Neutrons are more penetrating than $\gamma$-rays and they are a dangerous background 
source for most underground experiments. In dark matter experiments looking for direct interactions 
of WIMPs (Weakly Interacting Massive Particles), neutron 
background is critical because elastic interactions of fast neutrons give the very same signature 
as the signal, namely the WIMP elastic scattering off a nucleus. Similarly, $\gamma$-rays 
emitted in neutron inelastic scattering (n,n'$\gamma$) or capture (n,$\gamma$) (prompt or delayed) 
may mimic 
the signature of the neutrinoless $\beta\beta$ decay. Also unstable nuclei produced in
neutron-nucleus 
interactions are potential background sources in 
experiments looking for neutrinoless $\beta\beta$ decay. 

Hence, in order to predict and optimize the sensitivity of underground experiments, 
the background induced by neutrons must be precisely characterized.  
Monte Carlo simulations play a crucial role in this field, because:
\begin{enumerate} 
\item They can predict the neutron flux expected in an underground laboratory (as well as the 
energy spectrum) and the corresponding background rate due in a real experiment.
\item They can be used to study background suppression or rejection strategies, and to 
investigate requirements 
on the depth, the amount of active/passive shielding, the minimum veto efficiency, etc.,
for a given experiment.
\end{enumerate}
In this paper the Monte Carlo tools and methods that can be used to quantify neutron-induced 
background in underground experiments are reviewed. The actual validity of the Monte Carlo based 
calculation, and the achievable precision are also discussed. Finally, some considerations are 
presented about the most effective tools to reduce and suppress neutron background for underground 
physics experiments. The study has been focused on the experiments that are involved in 
the EU \textsc{Ilias} integrated activity~\cite{ilias}, and on the four underground 
Laboratories in the \textsc{Ilias} network, namely Gran Sasso (Italy), Modane (France), 
Canfranc (Spain) and Boulby 
(United Kingdom). The minimal depth of these underground sites is between 2.5 and 4~km w.~e. 
\section{Natural neutron sources in underground laboratories}

Neutrons in deep underground laboratories are produced in reactions initiated either by natural 
radioactivity or by cosmic rays. 

Neutron flux due to natural radioactivity (in the rock around the laboratory and/or in the materials 
of the experimental setup) is produced by: (1) spontaneous fission, the most important source being 
$^{238}$U, and (2)  ($\alpha$,n) interactions of $\alpha$'s from natural $\alpha$-emitters 
($E_{\alpha} < 10$~MeV) with light target nuclei. For heavy nuclei the ($\alpha$,n) cross 
section is suppressed by the Coulomb barrier. Neutrons from natural radioactivity 
have energy up to about 10~MeV. 

Neutrons are also produced in nuclear reactions (\emph{e.g.} 
muon-, photon- or hadron-induced spallation or disintegration) induced by cosmic ray muons. 
These reactions can be caused
by the muon itself (muon-induced spallation or negative muon capture) or by secondary
particles (photons and hadrons) generated in muon-induced cascades 
in the rock or in the materials of the 
experimental setup. The energy spectrum of these neutrons is substantially harder 
compared to neutrons from radioactivity.
Neutrons can be emitted with energies up to a few GeV. 

The relative importance of neutron-induced background coming from these two sources (natural radioactivity 
and cosmic rays) is very much dependent on the depth of the underground laboratory and on the 
specific experimental design, namely presence and thickness of passive shielding, muon vetoes, materials, 
etc. The Monte Carlo methods that are used to estimate the neutron flux in an underground laboratory and the 
neutron-induced background in a given detector are described in detail in 
sects.~\ref{sect3} and \ref{sect4}.

\section{Neutrons from natural radioactivity} \label{sect3}

\subsection{Ingredients for the calculation} \label{sect3_1}
The ``ingredients'' that are needed for the evaluation of the background induced in a given setup 
by neutrons from natural radioactivity are as follow:

\begin{enumerate}
\item The chemical composition of the source material (rock, detector component, etc.). The fractions of 
hydrogen and other light elements are particularly important because they affect in a 
substantial way the neutron production and/or propagation.
\item The contamination of the material in $^{238}$U, which undergoes spontaneous fission\footnote{Other 
naturally occurring isotopes are known to decay by spontaneous fission. Nevertheless, the branching ratios 
are negligible with respect to $^{238}$U.}, and in $\alpha$-emitters from the U and Th natural chains. 
The issue of secular equilibrium of radioactive chains is important both for the total 
neutron production rate and for the neutron energy spectrum, because the ($\alpha$,n) cross 
section is strongly dependent on energy. High-energy $\alpha$'s emitted by 
Po isotopes have in fact the highest probability to produce a neutron.
\item The nuclear parameters of interest. For spontaneous fission they are the multiplicity, the half-life 
and the relevant branching ratio; these data are available in nuclear databases 
for $^{238}$U 
and for other isotopes that decay by spontaneous fission. For the ($\alpha$,n) interactions 
it is necessary 
to know the interaction cross-section as a function of energy for all possible target isotopes 
that compose the material under investigation. It is also important to know the branching ratio for transitions
of the target nucleus to the excited states, since such a transition reduces the energy of the emitted neutron.
\item The propagation of neutrons from the production point to the 
sensitive volume, possibly through the external passive shielding, and the neutron detection process. 
The particle that is eventually detected is not necessarily the primary neutron, 
but may also be a secondary produced 
by neutron interactions (\emph{e.g.} a $\gamma$-ray from neutron capture or a recoiling nucleus). 
Propagation of neutrons through 
the experimental setup can be handled by Monte Carlo codes, such as 
\textsc{Geant4}~\cite{geant4-a,geant4-b} 
and \textsc{Mcnpx}~\cite{Mcnpx}. \textsc{Fluka} \cite{fluka} is also suited for this task but has
restrictions of not using point-wise cross-sections for neutron interactions and
of not generating individual nuclear recoils, acting as an event 
signature in dark matter experiments.
\end{enumerate}

The first three items are required to calculate the specific production rate of neutrons in the material under 
investigation, which is quoted in neutrons/~(cm$^{3}$~s) or in  neutrons/(kg~s). 
The last 
item is required to estimate the actual background rate (counts/s) in the detector; in this case the 
detector properties (\emph{e.g.} geometry, particle identification capabilities, etc.) have to 
be taken into account. 

The specific neutron production rate per unit volume for a given material can be calculated 
using the code \textsc{Sources}  
developed in the Los Alamos Laboratory~\cite{sources4a}. The software 
contains a database with tabulated ($\alpha$,n) cross-sections for several targets in the energy range 
of natural $\alpha$-emitters, energies of  $\alpha$'s for most $\alpha$-emitters and the main properties
(multiplicity, branching ratio, half-lives) for isotopes 
that undergo spontaneous fission. The original version of the \textsc{Sources 4A} code has been 
modified to extend the library of ($\alpha$,n) cross-sections and the
energy range of $\alpha$'s up to 10~MeV, in order 
to include all $\alpha$'s in the U and Th decay chains and more
target isotopes ~\cite{carson04}. 
Recently further improvements have been made to the code as part of the \textsc{Ilias} project
~\cite{lemrani06,vito07}.
More cross-sections have been added to the library as calculated using a dedicated code
\textsc{Empire 2.19}~\cite{empire}. The code has an advantage of being able to calculate branching
ratios for transitions to excited states. As specified above, this makes 
the calculated neutron energy spectrum softer, because part of the energy is 
carried away by $\gamma$-rays.
\subsection{Validation of the Monte Carlo codes}

For some target isotopes of interest experimental data on ($\alpha$,n) 
cross-section are scarce or totally unavailable. In this case, the cross-section has to be calculated 
using specific models. As discussed in sect.~\ref{sect3_1}, the cross-section database of the  
\textsc{Sources 4A} code has been extended using calculations from \textsc{Empire 2.19}. 
To prove the 
robustness of the calculations of neutron production rate it is necessary to validate the 
\textsc{Empire 2.19} predictions against experimental data.

\begin{figure}
\resizebox{0.5\textwidth}{!}{%
   \includegraphics{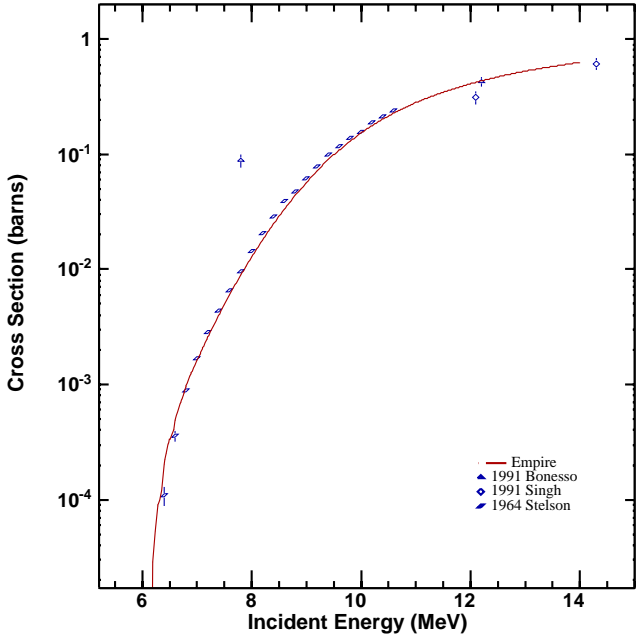}
   }
    \caption{Cross-section of ($\alpha$,n) reactions on $^{65}$Cu as a function of 
    $\alpha$ energy}
  \label{fig-alpha-cu}
\end{figure}

\begin{figure}
\resizebox{0.5\textwidth}{!}{%
   \includegraphics{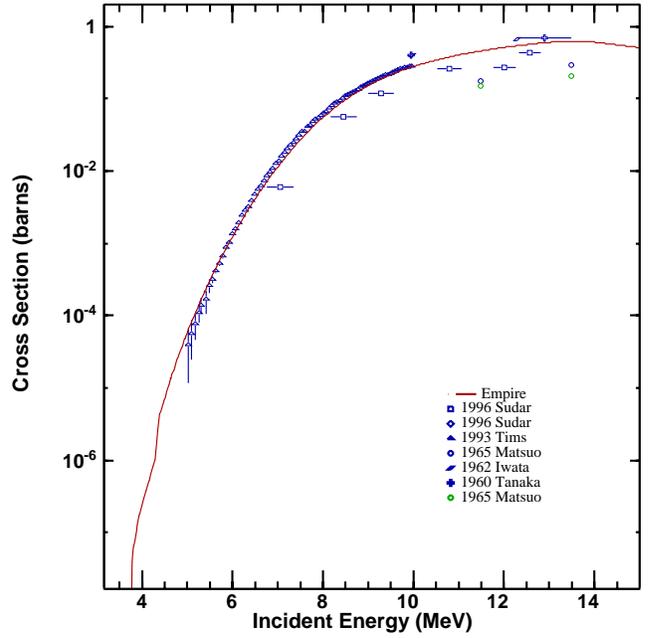}
   }
    \caption{Cross-section of ($\alpha$,n) reactions on $^{55}$Mn as a function of 
    $\alpha$ energy}
  \label{fig-alpha-mn}
\end{figure}

Figures~\ref{fig-alpha-cu} and \ref{fig-alpha-mn} show 
the ($\alpha$,n) cross-section versus energy as calculated with \textsc{Empire 2.19} for two
isotopes, in comparison with 
experimental data. The agreement is better than 20\% for the majority of experimental data
at energies below
10~MeV that are relevant for $\alpha$'s from radioactive isotopes. 
A large variety of tests were done to check the reliability of \textsc{Sources 4A} and
an agreement with experimental data was found for both neutron yields and spectra in
various materials (see, for instance, ref.~\cite{sources4a} for discussion).

The neutron-induced background rate in a given setup depends also on the transportation of 
neutrons from the production point in the source material itself. The 
two codes that are commonly used for neutron tracking are \textsc{Geant4} and \textsc{Mcnpx}. It 
has been demonstrated~\cite{lemrani06} that \textsc{Geant4} and \textsc{Mcnpx} are in good agreement 
for neutron tracking in the energy range of interest and in the materials that are commonly used for 
neutron shielding. As shown in ref.~\cite{lemrani06}, the deviation between the two codes 
after a mixed polyethylene/lead passive shielding is less than 50\% after a six-order of magnitude 
suppression of neutron flux. Such a suppression is required for high-sensitivity large-scale 
dark matter
and $\beta\beta$ decay detectors to reduce the neutron background from rock below
the experimental sensitivity level.  

Further confirmation of the \textsc{Geant4} ability to model neutron propagation and detection comes
from the agreement between simulations and measurements of fast ($\sim$~MeV) neutrons 
from a $^{252}$Cf source~\cite{tziaferi07}.

This demonstrates that the estimates of background due to neutrons 
from radioactivity in typical underground applications are reliable, with uncertainty being
less than 50\%.
\subsection{Results} \label{section3_3}

The most critical materials to be considered as neutron sources are mainly: (1) rock and concrete of 
the underground laboratories, that dominate the total neutron flux, 
and (2) materials that compose the 
internal detector parts (\emph{e.g.} stainless steel, copper, PMTs etc.), that become important when 
the external flux is suppressed by the (radio-pure) shielding.

\begin{figure}
\resizebox{0.5\textwidth}{!}{%
   \includegraphics{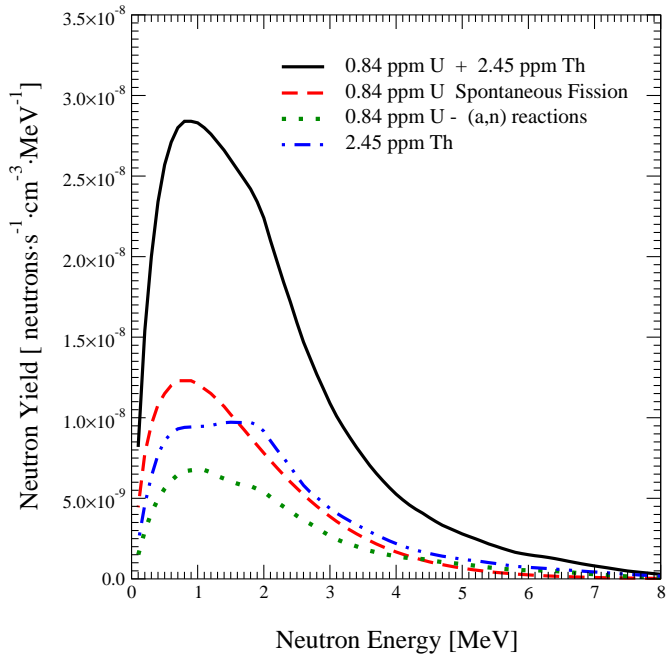}
   }
    \caption{Neutron energy spectra from uranium and thorium ($\alpha$,n) reactions,
    and $^{238}$U spontaneous fission in the
    LSM rock. The sum spectrum is shown by the solid curve.}
  \label{fig-nsp-rock}
\end{figure}

Figure~\ref{fig-nsp-rock} shows the neutron energy spectrum at production in rock around
the Modane Underground Laboratory (LSM). It is similar to that presented in ref.~\cite{lemrani06}
but has been obtained after recent improvements in the ($\alpha$,n) cross-sections 
calculations~\cite{vito07}. The rock composition and U/Th concentrations were taken from 
ref.~\cite{lemrani06}.  Separate contributions from uranium ($\alpha$,n) reactions, thorium
($\alpha$,n) reactions and $^{238}$U spontaneous fission are shown. 
Spectrum of neutrons from $^{238}$U spontaneous fission is described by 
the Watt function~\cite{watt}, with a peak energy of about 0.8~MeV and a mean energy
of about 1.7~MeV.
Spectra of neutrons from ($\alpha$,n) reactions are not much harder resulting in a
total energy spectrum (the sum of spontaneous fission and ($\alpha$,n) contributions)
with a mean energy of about 1.95~MeV. In general,
simulations with \textsc{Sources} taking into account the proper branching ratios for
transitions to excited states, give softer neutron spectra than reported in earlier
simulations and measurements (see, for instance, refs.~\cite{chazal98,arneodo99}).

\begin{figure}
\resizebox{0.5\textwidth}{!}{%
   \includegraphics{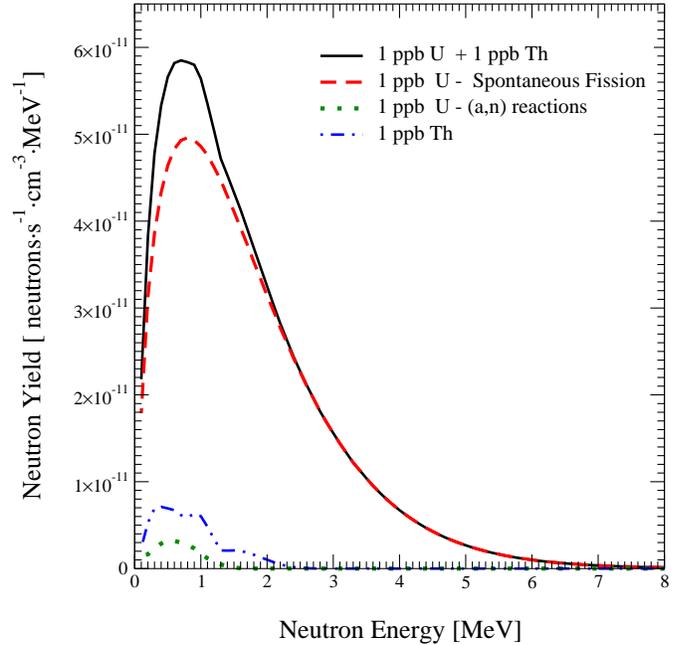}
   }
    \caption{Neutron energy spectra from uranium and thorium ($\alpha$,n) reactions,
    and $^{238}$U spontaneous fission in Cu. 
    Equal concentrations of 1 ppb were assumed for uranium and thorium.
    The sum spectrum is shown by the solid curve.}
  \label{fig-nsp-cu}
\end{figure}

The neutron energy spectrum from U and Th in copper is shown in Figure~\ref{fig-nsp-cu}.
Since the energy threshold for ($\alpha$,n) reactions in copper is about 7~MeV (due to high-$Z$
and Coulomb barrier), the main contribution comes from spontaneous fission of $^{238}$U.
Hence, for a given U/Th concentration, 
copper has an advantage over other materials containing lower-$Z$ isotopes (for instance,
stainless steel).

\begin{figure}
\resizebox{0.5\textwidth}{!}{%
   \includegraphics{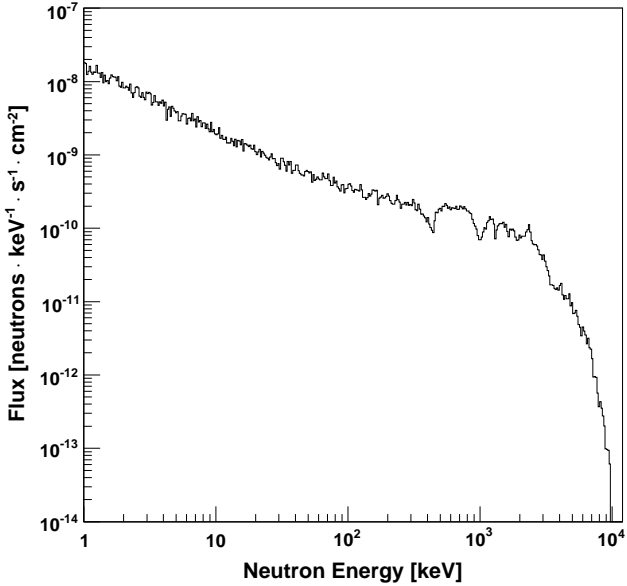}
   }
    \caption{Neutron energy spectrum from U and Th traces in the LSM rock. The spectrum
    is obtained by propagating neutrons from evenly distributed sources through rock to the
    laboratory walls using \textsc{Geant4}.}
  \label{fig-nsp-geant4}
\end{figure}

Figure~\ref{fig-nsp-geant4} shows the energy spectrum of
neutrons emerging from the LSM (Modane) rock into the laboratory. 
Neutron transport through the rock was carried
out using \textsc{Geant4} with an input spectrum from Figure~\ref{fig-nsp-rock}. 
Concrete walls were not included in these simulations and no back-scattering
of neutrons from the walls was taken into account. The latter effect increases
the neutron flux above 1 MeV by about 30\% \cite{lemrani06}.
Both total yield and spectral shape depend substantially on the chemical composition, in 
particular on the abundance of light elements and hydrogen. 
Hydrogen affects the thermalization and 
absorption of neutrons and hence suppresses the total flux. Only 1\% of hydrogen
reduces the neutron flux above 100~keV (1~MeV) by a factor of 4.7 (1.8)~\cite{lemrani06}.
The peaks and dips on the neutron spectrum are not of statistical origin but reflect the
shape of the cross-section of neutron interaction with rock. 
The precise knowledge of the chemical 
composition of the source material and possibly of its chemical homogeneity is crucial for 
reliable results, since it could be a dominant source of systematic uncertainty. 

Detailed comparison of the simulated neutron flux at LSM with the measurements were
presented in ref.~\cite{n-ed1}. Neutron yield was calculated with the \textsc{Sources 4A} code and 
neutrons were propagated using \textsc{Mcnpx}. Production of neutrons in rock and concrete
was taken into account. Experimental data were taken from ref.~\cite{chazal98}, corrected
using more accurate estimate of neutron detection efficiency \cite{lemrani05}.
It was found that the measured flux exceeds simulations by approximately a factor of 2
\cite{n-ed1}. We consider this as a reasonable agreement taking into account that:
(i) the conversion of the measured neutron rate in scintillator \cite{chazal98} into the neutron
flux and spectrum requires simulations and is not free from systematic uncertainties; 
(ii) the rock composition and U/Th concentrations are not known with high precision since
only a few samples of rock and concrete were measured; (iii) the average concentration 
of water in rock and concrete is difficult to measure whereas water (hydrogen) affects
significantly the neutron flux in the laboratory (see the discussion above).

Similar comparison between measurements and simulations was carried out 
for the Boulby rock (salt) \cite{tziaferi07}. The difference of about 50\% does not
exceed statistical and systematic uncertainties.

The typical fast neutron flux in underground laboratories is of the order of a few 
$10^{-6}$~neutrons/~(cm${^2}$ s). It depends 
on the parameters of the laboratory rock (composition and radiopurity) and is 
insensitive to the depth.

In low-background experiments, neutron flux from the rock has to be suppressed by 
means of dedicated 
thick passive shielding (polyethylene, water, etc.). A shielding thickness of about 
55-60~g/cm$^{2}$ of 
polyethylene (CH$_2$) is sufficient to suppress the external neutron flux by six orders of 
magnitude, 
ensuring a residual background of less than 1~event/tonne/year in direct dark matter 
experiments \cite{carson04}.
Similar suppression can be achieved by using 45-50~g/cm$^{2}$ of CH$_2$ together with
20-30 cm of lead placed between the rock and CH$_2$.
If the neutron passive shielding is in place, the dominant sources of neutron background 
from radioactivity are the detector 
components, giving $10^{-8}-10^{-10}$~neutrons/(cm${^2}$ s). For this reason, 
special care has to be taken of the radiopurity of the internal detector parts
(including shielding) and to 
the optimization of passive and active shielding. Further reduction of neutron background
due to radioactivity can be achieved by exploiting specific features of neutron
interactions as opposed to WIMP interactions, such as, for instance, multiple
nuclear recoils in the target, coincidences with active veto system, etc.
\section{Neutrons from cosmic-ray muons} \label{sect4}

The other important neutron source in underground laboratories are cosmic-ray muons and 
muon-induced cascades or showers. 
Neutrons produced by cosmic rays are substantially harder (extending up to GeV
energies) than those from radioactivity and are 
hence more difficult to suppress. Since the neutron flux is proportional to the residual muon rate  
at the underground site, 
this background can be reduced by going deeper and 
deeper. Nevertheless, it is important to estimate precisely the expected neutron flux induced 
by cosmic rays 
in the existing underground sites, and to optimize the suppression and rejection tools, in order 
to increase as much as possible the experimental sensitivity. 
\subsection{Ingredients for the calculation}
The calculation of the muon-induced neutron flux needs 
several inputs. The required ``ingredients'' are as follow:

\begin{enumerate}
\item The total muon flux at the underground site. Such a value is a specific parameter 
that characterizes 
each underground site, and is known experimentally. For instance, the residual muon 
flux at the Gran Sasso and Modane laboratories are about 
1.1~muon/(m$^{2}$ hour)~\cite{aglietta98,aglietta99,aglietta03} and 
0.23~muons/(m$^{2}$ hour)~\cite{rhode,berger89}, 
respectively\footnote{Muon flux is defined here as through a sphere with unit cross-sectional area.}.
The flux obviously decreases with depth. 
\item The energy spectrum and the angular distribution of muons. Energy and angle are in general correlated, 
depending on the particular surface profile above the underground laboratory. 
For laboratories located under 
mountains, the energy-angular distribution is determined by the structure and the 
orography of the region: muons 
come preferentially (and with softer energy spectrum) through the ``valleys'', where 
they meet the least rock overburden. 
Laboratories that are excavated in mines have a more homogeneous shielding profile. Information 
about energy and angular distribution of muons can be obtained by experimental measurements or, 
in the case they 
are not available or not reliable, by Monte Carlo simulation. Accurate measurements of muon
energy spectra underground are very difficult and one has to rely on simulations taking as an input
the muon energy spectrum at surface. Several fast dedicated codes, for instance 
\textsc{Music}~\cite{music}, are able to track muons through large distances in different materials, 
and to predict  
reliably the energy/angular distributions, provided the rock overburden 
profile of the site is accurately known. 
Another code \textsc{Musun}~\cite{musun} can be used to generate muons according to their 
angular and energy distribution (already obtained with \textsc{Music}) at the underground laboratory.
The average muon energy increases with the site depth: for the Gran Sasso laboratory it is about 270~GeV.
\item The code to track muons and their interactions, as well as 
production, propagation and possible detection of 
all secondaries, including neutrons. Monte Carlo codes that are commonly used for this purpose are 
\textsc{Geant4} and \textsc{Fluka}. 
\end{enumerate}

The first two items simply represent an input for the Monte Carlo tracking of primary muons. The third item 
is the most critical step, involving the modeling of the hadronic and electromagnetic interactions in a 
muon-induced particle shower. Usually a large number of secondary particles are produced 
within the shower, neutrons being a fraction of them. 
It is extremely important that all particles (including electrons and $\gamma$-rays) are 
produced in a Monte Carlo simulation and propagated through the whole setup. There 
is a possibility that neutrons are detected simultaneously with muons or 
other secondaries (interacting either in 
the sensitive detector or in other active volumes, such as an active veto), providing a very efficient tool 
for neutron background rejection.

\begin{figure}
\resizebox{0.5\textwidth}{!}{%
   \includegraphics{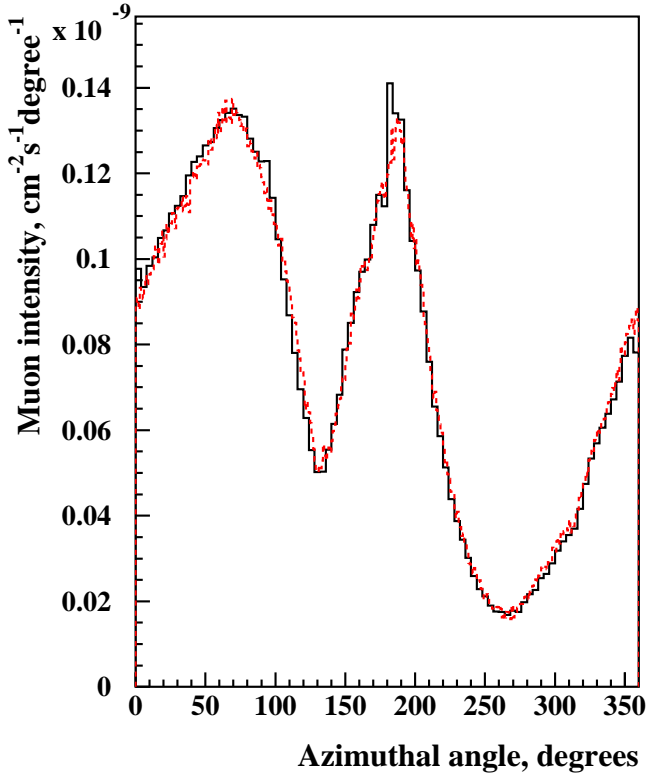}
   }
    \caption{Azimuthal angle distribution of single muon intensities, 
	measured by the LVD experiment in the Hall~A of the Gran Sasso Laboratory~\cite{aglietta99,aglietta03} 
	(black histogram), superimposed on the simulation results obtained with the  
	\textsc{Music}/\textsc{Musun} Monte Carlo codes (red dashed histogram).}
\label{musun-comp}
\end{figure}

Figure~\ref{musun-comp} shows the azimuthal angle distribution of single muon
intensities as measured by the LVD experiment at Gran Sasso 
(solid histogram)~\cite{aglietta99,aglietta03} together with the simulated one 
with the \textsc{Music}/\textsc{Musun} Monte Carlo codes. Muons were transported through 
the Gran Sasso rock using \textsc{Music}, their angular distributions and energy spectra were recorded
and used in \textsc{Musun}. The detector efficiencies have been taken into
account when plotting the distributions. The observed agreement proves the reliability of muon 
simulations, as described above. 

\subsection{Validation of the Monte Carlo codes}
Several output parameters/distributions can be derived from the Monte Carlo simulation 
(neutron energy and angular spectra, expected 
background rate in a real experiment, etc.). One possible output is the integral neutron 
yield, which is normally 
quoted in neutrons per muon per g/cm$^{2}$ of crossed target material. 
The integral neutron yield has to be compared 
to experimental data for specific targets, in order to validate and cross-check the simulations. 

Experimental measurements of muon-induced neutron yield are difficult because the flux 
is very low in deep underground laboratories. Only tonne-scale detectors are able to measure
the muon-induced neutron flux with reasonable accuracy. The neutron flux can be
artificially enhanced by placing a large amount of high-$A$ material around the detector.
Since the neutron yield increases with the atomic weight of the target material,
high-$A$ elements are the best targets for such measurements. Unfortunately, high-$A$
materials cannot be used for detecting neutrons. The best neutron detectors are based
on liquid scintillators that can be produced in large quantities at reasonable cost.
Also dark matter experiments are very good fast neutron detectors if accompanied by 
a muon detector, but require a large mass of target material.
Furthermore, the measurement is actually possible only at large distance from the 
primary muon track (otherwise the muon itself or secondary particles in muon shower are 
detected and single neutrons cannot be disentangled). 

The total neutron yield has been reported from different measurements
(see, for instance, refs.~\cite{musun,wang01} for a compilation and discussion of
experimental results). Although the authors converted their experimental results 
into the total yield per muon in the detecting material (\emph{e.g.} scintillator), 
the measured parameter was the
rate of thermal neutrons or, more precisely, the rate of $\gamma$-rays from neutron capture.
As in most experiments this rate was measured using liquid scintillator, a significant
part of fast neutrons was also detected through their moderation and capture in
scintillator.
The processes of neutron production, transport and detection are quite complicated and
the precise modeling of the detector geometry, hardware and software cuts and
all physical processes involved, is necessary for the accurate derivation of the neutron
yield. Unfortunately this has not been achieved so far for any experiment at large depth
underground.
Despite the lack of accurate simulations of specific experimental set-ups,
generic Monte Carlo results based of 
\textsc{Geant4} and \textsc{Fluka} are found to be in a reasonable agreement 
(within a factor of two)~\cite{musun,wang01,araujo05}
with the available data for low-$A$ materials (scintillator).
The overall agreement between \textsc{Geant4} and \textsc{Fluka} is quite
good for these materials. \textsc{Fluka} predicts the dominance of hadron-induced
spallation in neutron production for practically all targets, while 
\textsc{Geant4} favours nuclear disintegration by real
photons at almost all energies (except very low energies) and all materials.

While for low-$A$ targets the agreement between different codes and data
is reasonably good (certainly within a factor of two, if we neglect possible
systematic uncertainties due to the absence of precise modeling
of the experimental setups), some experiments with heavy targets showed
much larger neutron yield than expected. Differential 
cross-section of neutron production in thin targets (graphite, copper, lead) has been 
measured by the NA55 experiment at CERN for 190-GeV muons~\cite{chazal}. 
The thin-target configuration 
does not correspond to the real situation in underground laboratories, where showers 
can develop through large 
thickness of rock but can be modeled more accurately. 
Experimental data have been compared to Monte Carlo simulations 
performed with \textsc{Geant4} and 
\textsc{Fluka} in ref.~\cite{araujo05}.
While \textsc{Geant4} and \textsc{Fluka} agree with each other within a factor of two, 
both codes underestimate significantly 
the neutron production as measured by NA55, especially
for copper and lead.
Yet again, a complete simulation of the NA55 
detector has not been carried out by the authors of the original work 
and it is difficult to draw definite conclusions, because some detector-specific aspects 
(geometry, response) could have been neglected in ref.~\cite{araujo05}.

There are other experimental data available for neutron yield in lead, obtained 
in deep underground laboratories.
These data are old and controversial~\cite{bergamasco,gorshkov}, 
but also indicate higher neutron production in lead than expected from
modern Monte Carlo simulations. 
The issue of neutron production in lead is extremely important to  
underground experiments, because 
lead is commonly used as shielding material.
More experiments are planned as part of the \textsc{Ilias} activities
using underground dark matter detectors or massive 
veto systems, accompanied by detailed Monte Carlo simulations 
to be used for data interpretation. 

Preliminary results from the measurements of muon-induced neutrons
at the Boulby Underground Laboratory have recently been reported
\cite{vak07}. This experiment improves on previous, even larger scale detectors,
since full 3D Monte Carlo of the set-up has been carried out using \textsc{Geant4}.
Preliminary results suggest that \textsc{Geant4} overproduces neutrons in the
materials surrounding the detector (mainly rock and lead) by about 80\%
compared to the measured muon-induced neutron rate. This is not consistent with
the deficit of simulated neutrons discussed in ref.~\cite{musun,wang01,araujo05}
(see also references therein).
Detailed comparison of muon-induced neutron background as simulated
with \textsc{Geant4} and \textsc{Fluka} has been carried out in ref.~\cite{araujo05}.
The agreement within 20\% has been found for neutron spectra in an
underground laboratory located in salt rock (Boulby). Special case of
muon-induced neutron background in a xenon-based dark matter experiment
has also been investigated. \textsc{Geant4} and \textsc{Fluka} are consistent in predicting
the rate of events caused by this background to within 20\% (the difference
does not exceed statistical uncertainty of simulations). The observed inconsistency
between measured and simulated neutron yields makes the predictions of muon-induced
neutron rate in various detectors uncertain by about a factor of two.
\subsection{Results}

Neutron yield is dependent on the target material, ranging from a few 
$10^{-4}$ neutrons/muon/(g/cm$^{2}$) 
for light materials (water, graphite, liquid scintillator) up to a few
$10^{-3}$ neutrons/muon/(g/cm$^{2}$) for 
high-$A$ elements, such as gold and lead. 
Monte Carlo simulations have been used to derive a scaling law for neutron yield $Y$ versus 
the atomic mass $A$, namely $Y \propto \ A^{0.8}$.

Since high-$A$ targets, such as lead that is used for passive $\gamma$-ray shielding, 
have higher neutron yield, they behave 
like a neutron source under muon irradiation. For this reason it is necessary to 
include an internal low-$A$ shielding (preferably with large amount of hydrogen, for instance,
polyethylene or water) between the lead shielding and the sensitive detector, in 
order to reduce the muon-induced neutron background. 

Neutrons produced in the rock, lead shielding or other 
materials of the experimental setup can be identified and tagged if the primary muon or other secondaries interact 
in the detector or in an active veto surrounding the detector. Indeed only 
neutrons that are not accompanied by 
other particles in the detector or in active vetoes represent a background 
for underground experiments. Full discussion of these issues with respect
to dark matter searches can be found
in ref.~\cite{araujo05}.

The typical flux of fast neutrons induced by cosmic ray muons in an underground 
laboratory with a rock coverage of 
3000~m~w.~e. (as Gran Sasso or Boulby) is about $10^{-9}$~neutrons/(cm${^2}$ s), 
\emph{i.e.} three orders of magnitude 
smaller than the neutron flux produced by radioactivity. The flux of muon-induced neutrons 
is strongly dependent on 
the depth of the underground site, that affects the 
total muon flux and spectrum, as well as on the composition of the 
rock (density and average atomic mass $\langle A \rangle$). 

\begin{figure}
\resizebox{0.5\textwidth}{!}{%
   \includegraphics{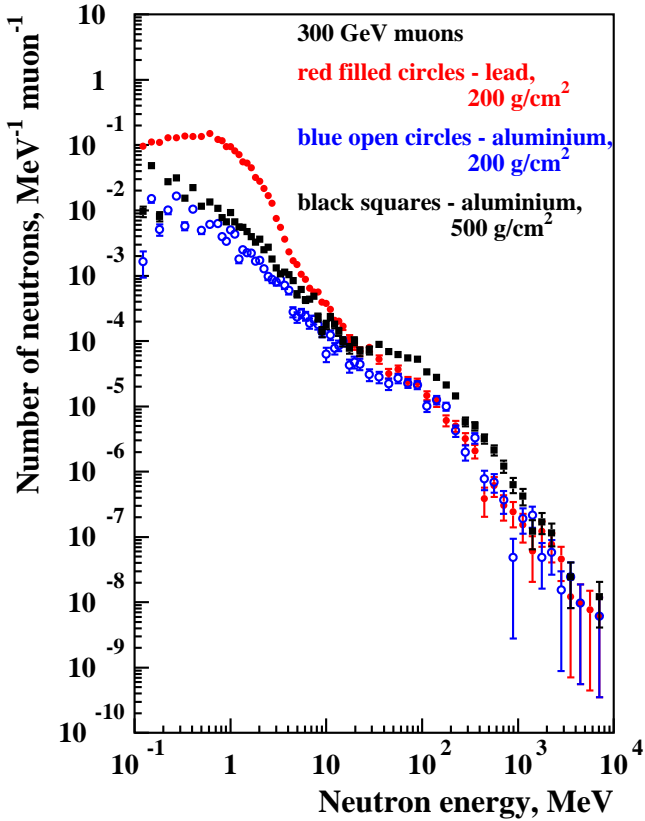}
   }
    \caption{Neutron energy spectra from 300~GeV muons in lead (thickness
    of 200~g/cm$^2$ -- filled circles) and aluminium (thickness
    of 200~g/cm$^2$ -- open circles and
    of 500~g/cm$^2$ -- filled squares)
    as calculated with \textsc{Fluka} (version of 2003). }
  \label{fig-nsp-mu}
\end{figure}

Energy spectrum of muon-induced neutrons is substantially harder than 
that from fission or ($\alpha$,n) reactions. 
Figure~\ref{fig-nsp-mu} shows the neutron spectra in lead and aluminium 
(mean atomic weight similar to rock) from 300~GeV muons. The spectra were obtained by
transporting a beam of muons (from `vacuum') through a certain 
thickness of material and recording the secondary neutron spectrum at the boundary
between the material and the vacuum beyond it.
Two important conclusions can be derived from this figure. Firstly, 200~g/cm$^2$ thickness of
target is too small for a neutron flux to be in equilibrium with the muon flux. Neutron flux increases
as the thickness of aluminium increases to 500~g/cm$^2$. For this reason, when interpreting the
data from experiments with relatively thin targets, a proper account should be taken of the 
development of muon-induced cascades, which is impossible to achieve without full Monte Carlo
(see also ref.~\cite{araujo05} for discussion).
Secondly, the neutron energy 
spectrum depends strongly on the target material. All enhancement of the 
neutron production in lead occurs at neutron energies below 20~MeV. The flux above
20~MeV is practically material independent. Hence, high-$A$ targets give higher neutron yield 
than low-$A$ ones, but with softer energy spectrum. Similar conclusion was derived also
in refs.~\cite{carson04,musun,wang01,araujo05}.

\begin{figure}
\resizebox{0.5\textwidth}{!}{%
   \includegraphics{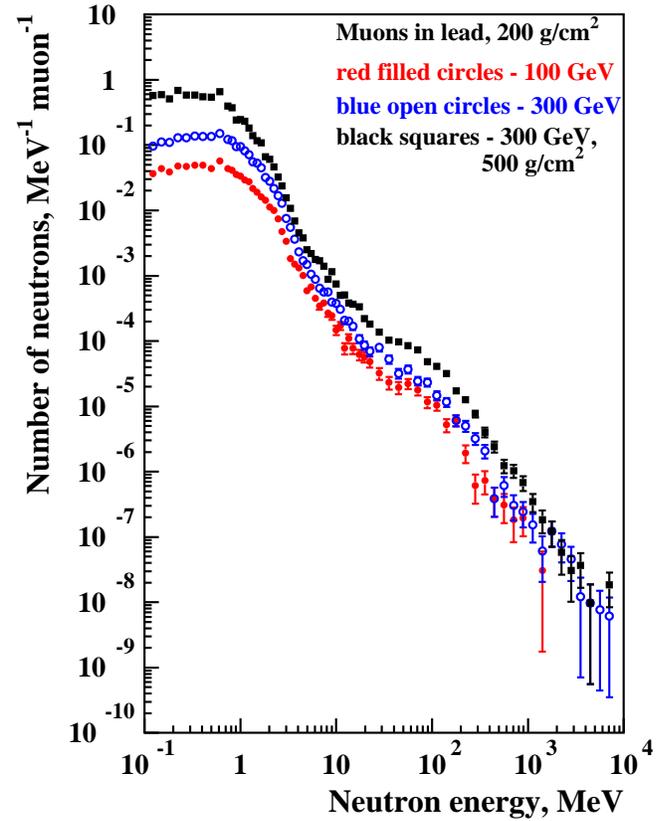}
   }
    \caption{Neutron energy spectra in lead from 100~GeV muons (thickness
    of 200~g/cm$^2$ -- filled circles) and 300~GeV muons (thickness
    of 200~g/cm$^2$ -- open circles; 500~g/cm$^2$ -- filled squares)
    as calculated with \textsc{Fluka} (version of 2003). }
  \label{fig-nsp-mulead}
\end{figure}

Figure~\ref{fig-nsp-mulead} shows the neutron energy spectra produced by muons with
energies 100~GeV and 300~GeV in lead with thickness of 200~g/cm$^2$ and 500~g/cm$^2$.
The spectra were obtained in the same way as above.

The figure demonstrates that the neutron flux increases with the increase of lead thickness
from 200~g/cm$^2$ to 500~g/cm$^2$ but the enhancement is more pronounced at energies
below 1-2~MeV. This is probably due to the secondary low-energy neutron production by higher
energy neutrons. The neutron flux also increases with muon energy. The change
in the neutron spectrum with muon energy is higher at neutron energies below a few MeV.
\section{Background suppression strategies}

A number of tools (both hardware and software) is available to suppress or tag 
neutron-induced background in underground physics experiments.
To reduce background due to neutrons produced by natural radioactivity 
in the rock around the laboratory and in 
the detector materials, three ways have to be considered:

\begin{itemize}
\item A passive neutron shielding (made of water, polyethylene or other hydrogenous 
material) to suppress the external neutron flux due to the rock radioactivity, which is of 
the order of a few $10^{-6}$~neutrons/(cm${^2}$ s). 
It has been shown~\cite{carson04,lemrani06} that 50-60 g/cm$^{2}$ of polyethylene 
(possibly with lead) can 
reduce the neutron flux from the rock by six orders of magnitude. 
\item Special attention has to be paid to material selection for radiopurity. 
Neutron background in the 
detector (in the presence of a thick passive shielding, as indicated above) 
is mainly originated in the materials used in detector or shielding construction. 
This may eventually become the limiting background source.
\item An active veto system can be installed just around the detector to tag
neutrons produced in detector components~\cite{carson05,pfs05}. 
A neutron can give a signal in - say - 
a dark matter detector, be scattered, slowed down and detected through its capture
in the active veto system. Simultaneous detection of neutrons and/or $\gamma$-rays 
by the main target and veto can also be used to tag the background in 
$\beta\beta$ decay experiments.
\end{itemize}

Similarly, several methods have been proposed for the suppression or tagging 
of high-energy neutrons produced 
by muon interactions in the rock surrounding the detector and in the detector 
materials themselves, especially 
in the high-$A$ $\gamma$-ray shielding. 

\begin{itemize}
\item It is necessary to have an internal low-$A$ neutron shielding, 
in order to shield the sensitive detector 
from neutrons produced by muon interactions in the rock and in the $\gamma$-ray shielding. 
A two- or three-layer shielding 
(external low-$A$ (optional) + high-$A$ + low-$A$) could be hence optimal for most applications.
\item Since in muon-induced showers neutrons are accompanied by a large number of other secondaries, rejection 
tools based on veto or self-veto are extremely efficient. There is a large probability 
that a neutron is detected simultaneously with the primary muon or with other 
secondaries in the detector itself (self-veto) 
or in an active veto system~\cite{carson04,araujo05}. It was 
shown~\cite{carson04,araujo05} that in a large
(250~kg) xenon-based dark matter detector, less than 5\% of events with nuclear recoils
are actually single nuclear recoils in the energy range of interest for dark matter searches. 
The remaining events contain energy deposition from
nuclear recoils and muons or other secondaries produced by muons
and hence do not mimic WIMP interactions. Similar result is expected for other
detectors.
\item The detector design has to be optimized, especially concerning the mass and 
position of high-$A$ materials, since they may behave as neutron sources under 
muon irradiation. 
\end{itemize}

Muon-induced neutron background depends strongly 
on the design of the experiment~\cite{pandola07}. 
In particular, properties and placement of the materials close to the 
detector are extremely important, because they affect the muon showering and the propagation of secondaries.  
The shielding design has to be optimized as a compromise 
between the suppression of external radiation, the reduction of muon-induced background 
produced within the setup and the overall cost of the setup. 
Detector-specific and site-specific accurate Monte Carlo simulations are indeed 
required to achieve this goal.

Monte Carlo studies demonstrate that, with the choice of a suitable 
shielding and taking advantage from  
the tagging and rejection tools listed above, neutron-induced background 
does not represent a limitation 
for the existing and next-generation underground experiments at the depth 
of 2.5-4~km~w.~e., namely 
the depth of the European laboratories involved in \textsc{Ilias}. 
In particular, the neutron background rate  
can be reduced down to a few events/tonne/year for direct dark matter experiments 
(corresponding to a potential cross-section sensitivity of $10^{-10}$~pb to spin-independent
WIMP-nucleon interactions) \cite{araujo05}
and to $10^{-3}$~events/(keV kg year) 
for neutrinoless $\beta\beta$ decay experiments \cite{pandola07}.
\section{Conclusions}

Neutrons produced in underground laboratories either by natural radioactivity or 
by muons are an 
important background source for underground experiments looking for rare events 
(dark matter, neutrinoless $\beta\beta$ decay, low-energy neutrinos, etc.). 
Monte Carlo simulations play a crucial 
role for the estimate of the residual neutron-induced 
background and for the optimization of the rejection strategies. 
It is necessary that simulations are as detailed 
as possible, including detector-related effects (like geometry and response).

The neutron flux due to radioactivity can be reliably predicted, 
using the combination of computer codes, such as \textsc{Sources 4A} for 
the estimate of the neutron yield and \textsc{Geant4}/\textsc{Mcnpx} for the 
description of neutron propagation and detection. 
The main systematic uncertainties are typically related to the  
knowledge of the chemical composition 
(and homogeneity) of the target material, and possibly to the ($\alpha$,n) cross-sections. 
From the comparison between simulated neutron fluxes and measurements
in underground laboratories (sect.~\ref{section3_3}), 
we can estimate the above uncertainty as about a factor of two.
Since neutron flux 
from the laboratory rocks can be effectively shielded, neutrons emitted by 
detector components represent the 
main contribution to the total background, which may eventually limit the sensitivity. 
This background can be 
minimized by means of a careful material selection for radiopurity.

Neutron background induced by muon interactions is strongly dependent on 
the depth of the laboratory and on 
the experimental design, especially concerning position of high-$A$ 
materials. In most 
cases neutrons produced by muon interactions in lead $\gamma$-ray shielding 
have to be suppressed by 
means of an internal neutron shielding made of hydrogenous materials. 
Since muon-induced showers 
contain a large number of particles, neutrons in the detector are often 
accompanied by other secondaries. For this reason, 
rejection methods based on simultaneous interactions of other particles in the 
detector itself (self-veto) 
or in an external veto are extremely efficient. 
To estimate reliably this effect, all particles produced 
in the showers have to be created and propagated in the Monte Carlo simulation. 
The two most-commonly 
used codes to describe the propagation of muon-initiated showers 
(\textsc{Geant4} and \textsc{Fluka}) 
agree within a factor of two for the integral neutron yield. 
They both agree with measurements performed 
for low-$A$ target materials (\emph{e.g.} liquid scintillators), while seems to under-produce 
neutrons for 
heavier materials, such as lead (although experimental data are controversial 
and sometimes inconsistent). New 
experimental data are being collected in underground laboratories, 
that have to be considered together with a detailed Monte Carlo simulation 
accounting for detector-related effects.

Given the available suppression tools, neutron-induced background does not limit 
the sensitivity 
of existing and next-generation underground experiments 
at the depth of the \textsc{Ilias} European laboratories, \emph{i.e.} between 
2.5 and 4~km~w.~e. 
\section{Acknowledgments}
This work has been supported by the \textsc{Ilias} integrating activity (Contract 
No. RII3-CT-2004-506222) as part of the EU FP6 programme in 
Astroparticle Physics (including the Ph.D. research of V.~Tomasello).
We would like to thank all our colleagues involved in the \textsc{Ilias} activities, 
especially concerning Monte Carlo simulations (\textsc{N3} and \textsc{JRA1}), in
particular Drs. M.~Robinson, R.~Lemrani, M.~J.~Carson, H.~Ara\'ujo and Mr. H.~Chagani. 
We are grateful to the organizers of the \textsc{Ilias} meeting at Chamb\'ery for giving us
an opportunity to present this review. 

We dedicate this work to the memory of our friend and colleague Nicola Ferrari, 
who prematurely passed away in July 2006.

%

%

%
%

\end{document}